# Phase-continuous comparison of three all-optical time scales over 20 days

Dahyeon Lee[1], Kyungtae Kim[1], Zoey Z. Hu[1], Ben Lewis[1], William Warfield[1], Kai Zhou[1], Alejandra L. Collopy[2], Jeffrey A. Sherman[2], Abijith S. Kowligy[3], Parth B. Patel[3], Jonathan D. Roslund[3], Arman Cingöz[3], Martin M. Boyd[3], Jun Ye[1]

[1]JILA, National Institute of Standards and Technology and University of Colorado, Department of Physics, University of Colorado, Boulder, CO 80309, USA

[2]National Institute of Standards and Technology, Boulder, CO 80305, USA

[3]Vector Atomic, Inc., an IonQ company, Pleasanton, CA 94566, USA

Optical frequency standards have progressed rapidly over the past two decades, leading to the anticipated redefinition of the SI second by an optical frequency. However, time scales have not yet significantly improved despite this development because they are still fully reliant on rf flywheel oscillators, mostly hydrogen masers, which impose a performance limit related to incompletely sampled noise known as the Dick effect. To best benefit from the exceptional stability and accuracy of optical frequency standards, time scales must employ optical flywheels with orders-of-magnitude better short-term ($<10^4$ s) stability than masers. Here, we introduce three optical flywheel oscillators (two cryogenic silicon cavities and one iodine optical clock) with superior short-term stability than hydrogen masers and long-term stability on par with masers. Steering each optical flywheel with a high-uptime Sr optical frequency standard generates three parallel all-optical time scales with continuous operation over >20 days. When compared with each other, these all-optical time scales achieve $<10^{-16}$ relative instability after just a few days of averaging. During typical steering gaps of ~6 hours, the accumulated time difference is ~20 ps, leading to the total time difference of <100 ps over the full measurement period. With the proliferation of long-distance optical fiber links and commercialization of optical flywheels and frequency standards, we anticipate all-optical time scales to be the future of timekeeping.

## Introduction

Optical frequency standards provide orders-of-magnitude improved frequency stability and potential accuracy compared to the rf frequency standards that currently realize the SI second[1,2]. The superior performance of optical frequency standards promises more precise timekeeping, motivating the anticipated redefinition of the SI second in terms of an optical frequency[3,4].

However, there has not been a commensurate performance improvement in time scales such as Coordinated Universal Time (UTC) and its local realizations by national metrology institutes. This is partly because time scales are still exclusively realized with rf flywheel oscillators, mostly hydrogen masers, offering robustness but lacking the stability performance of optical frequency standards[5]. Hydrogen masers exhibit at least a thousand times worse frequency stability than optical frequency standards, which means even a modest amount of steering downtime is enough to introduce a strong Dick noise limit well above the native performance of the optical frequency standard[6–8]. As a result, maser-based time scales, even when steered by optical frequency standards, require weeks of averaging to reach the frequency stability that an optical frequency standard would reach in 1 s.

This severe underutilization of resources can be remedied if a suitable optical flywheel oscillator is used to keep time entirely in the optical domain, forming an all-optical time scale. Optical flywheels offer enhanced short-term ($<10^4$ s) stability compared to masers, resulting in a correspondingly lower Dick noise limit. Cryogenic silicon cavities are realistic candidates for time scale flywheels because they simultaneously achieve short-term fractional frequency stability of mid-$10^{-17}$, ~1000 times lower frequency drift than room-temperature Ultra-low Expansion glass (ULE) cavities[9], and long-term stability on par with hydrogen masers. The possibility of using a cryogenic silicon cavity for an all-optical time scale was explored in Ref. 10 with $48 \pm 94$ ps estimated time error after 34 days of operation. Other candidates for optical flywheels include a molecular iodine optical clock[11] and Ca beam clock[12], both offering improved short-term stability than hydrogen masers. The most challenging requirements on the optical flywheel are reliability and long-term frequency stability, particularly at averaging times corresponding to the typical duration of gaps in the frequency standard operation. Because these requirements are not met by most optical flywheel oscillators, efforts to incorporate optical frequency standards to time scales so far have been limited to steering of hydrogen masers[8,13–21].

As a demonstration of the feasibility and performance advantages of using optical flywheels for time scales, we realize three parallel all-optical time scales using two state-of-the-art cryogenic silicon cavities and one commercial iodine optical clock, each steered with a $^{87}$Sr optical lattice frequency standard. These three time scales are phase-continuously compared with each other for more than 20 days. Because of the superior short-term stability of the optical flywheels and the high uptime (~70%) of the Sr frequency standard, our all-optical time scales achieve sub-$10^{-16}$ fractional frequency instability after just a few days of averaging, significantly faster than the weeks it would take a maser-based hybrid optical time scale to reach the same performance[8]. At the end of the measurement campaign, the relative time differences between the time scales are <100 ps, much smaller than the nanoseconds typical of conventional time scales over a similar duration. In addition, the all-optical time scales are downconverted to the rf domain with the aid of optical frequency combs; the fidelity of the optical-to-rf conversion process is verified in terms of both stability and accuracy.

## Experiment and results

An important consideration for pursuing all-optical time scales is the reliability of optical flywheels (lasers). A time scale flywheel must continuously produce a clock signal, ideally indefinitely. We employ 1.5 μm fiber lasers that stay locked to cryogenic silicon cavities for months at a time if unperturbed by electrical power disturbances. The iodine optical clock is a commercial unit that operates for extended periods of time, even outside the lab[22]. Robust, continuously locked optical frequency combs are now commercially available from multiple vendors[23–25]. The robustness of modern optical frequency standards is also improving[26–28], with commercialization efforts underway. The reliability of these time scale sub-systems makes the possibility of implementing all-optical time scales realistic.

Figure 1(a) shows a conceptual diagram of the experiment. Using a Sr optical frequency standard, we measure and steer the rates of three independent optical flywheels. The resultant time scales share the state-of-the-art optical frequency standard but are otherwise independent; therefore, each time scale's relative performance is determined only by its steering and intrinsic noise of the optical flywheels. Each all-optical time scale is paired with an optical frequency comb to convert it to the rf domain. The time scales are phase-continuously compared with each other in both optical and rf domains.

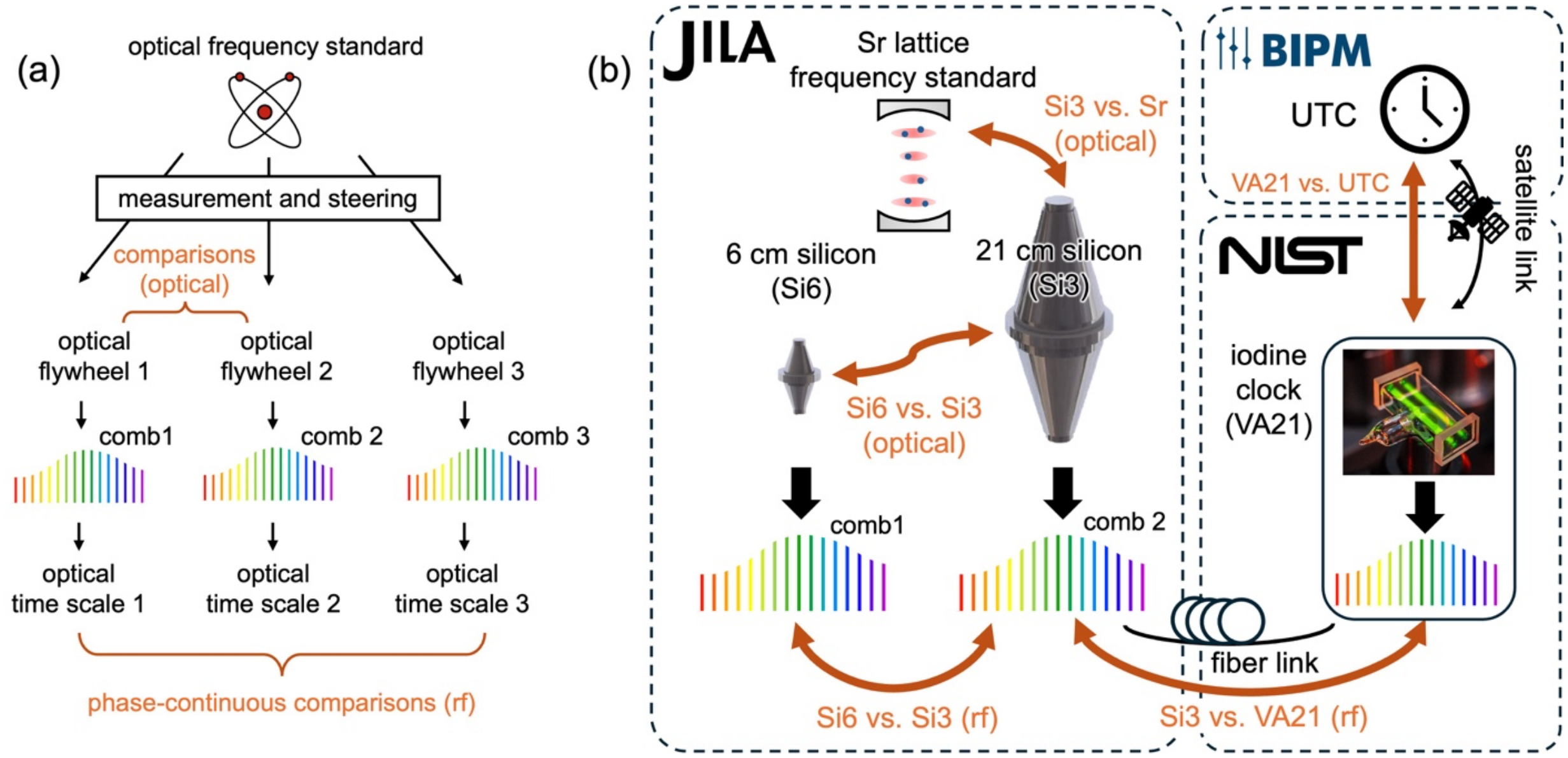

**Figure 1** Schematic of the experimental setup. (a) Three all-optical time scales based on three independent optical flywheels are generated and continuously compared with each other. The optical frequency standard measures and steers the frequencies of the optical flywheels. Each optical flywheel frequency is downconverted to the rf domain with an optical frequency comb. (b) Details of the experimental setup. Orange arrows indicate the physical measurements performed. These measurements are used to derive additional comparisons, for example, Si6 vs. Sr and VA21 vs. Sr. All measurements are made continuously without interruptions so that the phase (time) information is preserved, except for the ones involving the Sr frequency standard. The microwave signal link between JILA and NIST is through an actively phase-stabilized rf-over-fiber transmission and the one between NIST and BIPM is through a satellite.

Figure 1(b) shows implementation details of the all-optical time scales. For brevity, we use abbreviations “Si3” for the 21 cm silicon cavity[29,30], “Si6” for the 6 cm silicon cavity[9,31], “VA21” for the molecular iodine clock[22] (Vector Atomic EG-30), and “Sr lattice” for the Sr optical frequency standard[30,32,33]. Si3 serves both as a local oscillator for the Sr lattice and a time scale flywheel. An actively phase-stabilized rf-over-fiber transmission provides a microwave signal link between JILA and NIST[34,35]. The physical measurements performed are highlighted in orange arrows. These measurements are appropriately combined to derive additional comparisons, for example, between VA21 and the Sr lattice. The setup is compatible with both real-time and post-processed time scale generation, and the latter approach is used to present the data here.

By a similar chain of oscillator measurements, our time scales are connected to UTC through satellite-mediated time transfer links, paving the way for the Sr lattice to contribute to the rate adjustments of International Atomic Time (TAI). Furthermore, this enables the future integration of the optical flywheel oscillators to the clock ensemble Echelle Atomique Libre (EAL)[36]. Figure 2(a) shows a 7-month-long measurement of the frequencies of Si3, VA21 (reported to BIPM as clock 1442021), and NIST maser ST8 (BIPM: 1412108), which also currently serves as the local oscillator to NIST cesium fountains NIST-F3 and NIST-F4. The Sr lattice operated with 51% uptime over this period. We mostly operate our optical frequency standard overnight to accommodate other experiments during the day, so the uptime would be higher for a fully dedicated system (see Extended Data Fig. 2(b)). The short-term frequency fluctuation of Si3 is shown in the inset with a y-axis magnification of 1000 times. The long-term frequency predictability of Si3 is comparable to that of maser ST8, which justifies the silicon cavity's usefulness as a time scale flywheel.

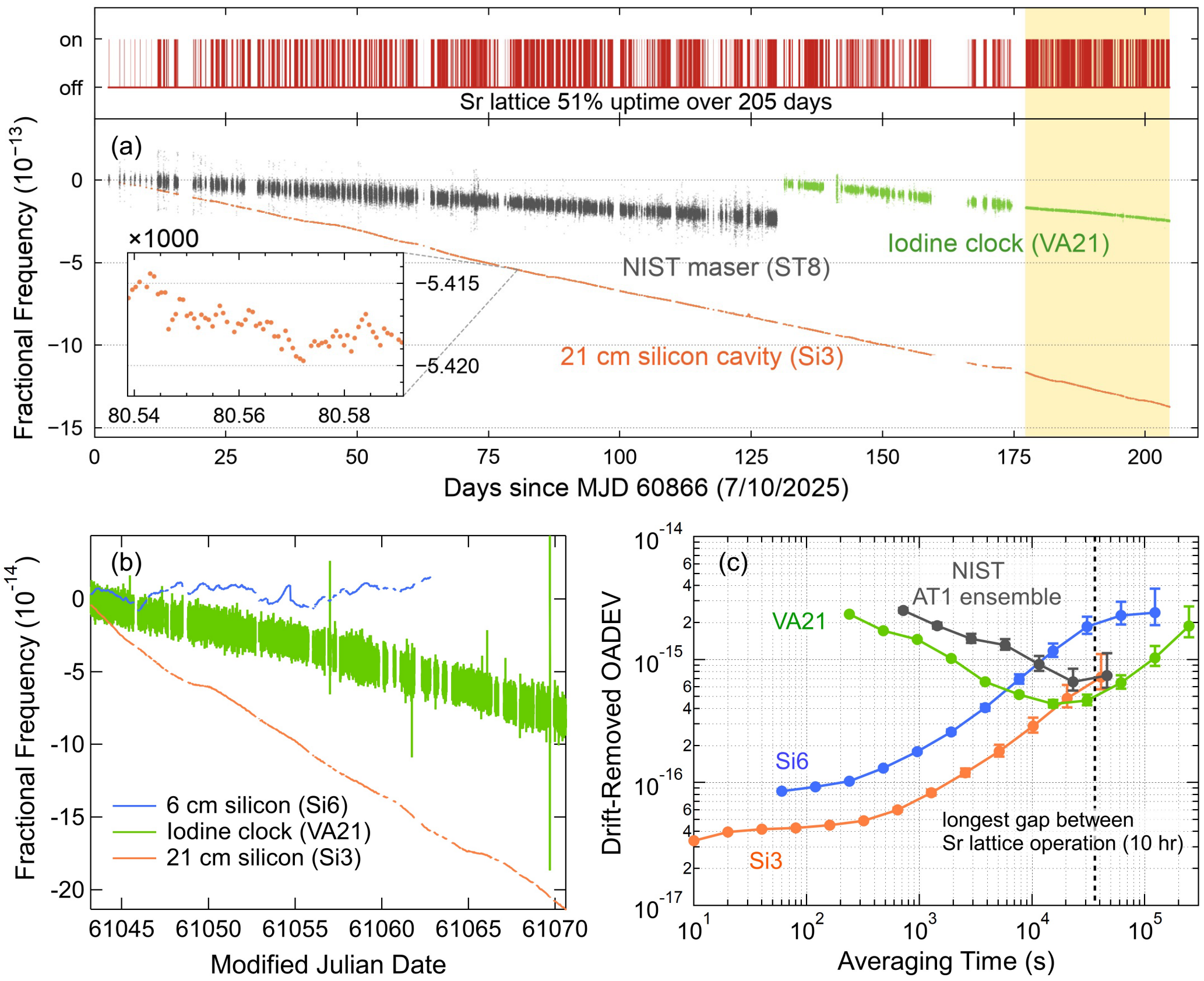

**Figure 2** High-uptime operation of the Sr optical frequency standard. (a) Frequency record of a cryogenic silicon cavity (Si3), optical iodine clock (VA21), and hydrogen maser (ST8) over 7 months, measured with the Sr frequency standard with 51% uptime. All-optical time scales were generated in the region highlighted in yellow. During this period, Si3 and VA21 had 100% uptime and the measurement noise floor for VA21 was improved. One-minute averaging interval is used for all traces. (b) In January 2026, the frequencies of the three flywheels were measured with the Sr frequency standard with ~70% uptime. This record was used for steering the flywheels to generate all-optical time scales. Si3 and Si6 were measured in the optical domain (194 THz) and VA21 was measured in the rf domain (5 MHz). The sharp peaks in the VA21 trace are from fiber link instability. (c) Drift-removed overlapping Allan deviation of the flywheel oscillators.

We generated all-optical time scales for 27 days during MJD 61043-61070 (January 2026), highlighted in Fig. 2(a), with a higher Sr lattice uptime of 71%. Figure 2(b) shows the frequencies of all three optical flywheels relative to the Sr lattice. The 20-day-long Si6 measurement was terminated when a brief power outage stopped the cryostat after 1 year of continuous operation. Si6 does not show a discernable drift over this period. VA21 shows a predictable drift over multiple days of low-$10^{-15}$ per day and a typical short-term stability of $3 \times 10^{-14}$ at 1 s. Si3 is the most stable over a few hours, which makes it the best flywheel given that the steering gaps are always less than 10 hours in our case. The intrinsic noise characteristics of the flywheels are captured by the drift-removed overlapping Allan deviation of Fig. 2(c). The dashed vertical line marks the duration of the longest gap between Sr lattice operations in the entire measurement campaign. Since the Allan deviations of Si3 and VA21 at shorter averaging times are comparable to or better than the NIST maser ensemble AT1 that forms UTC(NIST), we expect all-optical time scales based on these optical flywheels to perform better than UTC(NIST) at all averaging times.

Effective steering of a flywheel requires modeling and prediction of future clock behavior. Here, we take the simplest approach of compensating only the predictable linear drift of the silicon cavities in the absence of information from the Sr lattice. Specifically, the algorithm for steering the silicon cavities is as follows: when the Sr lattice is running, the average measured frequency over the past 1 minute is used to correct the current frequency. The steering frequency is updated every minute. The choice of the one-minute averaging interval corresponds to the minimum required averaging time for the measurement uncertainty to reach the silicon cavity's flicker frequency noise floor. When the Sr lattice stops running, the predicted frequency is linearly extrapolated from the last measured value. For Si3, the slope of the extrapolation is taken as the average drift rate over the past 10 hours. Figure 3(a) shows the measured frequency of Si3 (red) and the correction applied to it (blue). The correction applied while the Sr lattice is running cannot be seen at the plotted scale because it almost perfectly overlaps with the Sr measurement. When the Sr lattice is offline, the algorithm corrects for the estimated frequency drift until the next measurement becomes available. For steering Si6, shown in Fig. 3(b), the linear drift is modeled as zero because of poor predictability. As a result, the applied correction stays constant until the next measurement becomes available. For steering VA21 (Fig. 3(c)), we use a Kalman filter algorithm similar to the approach used for steering a hydrogen maser with an optical frequency standard[8,13,17–19].

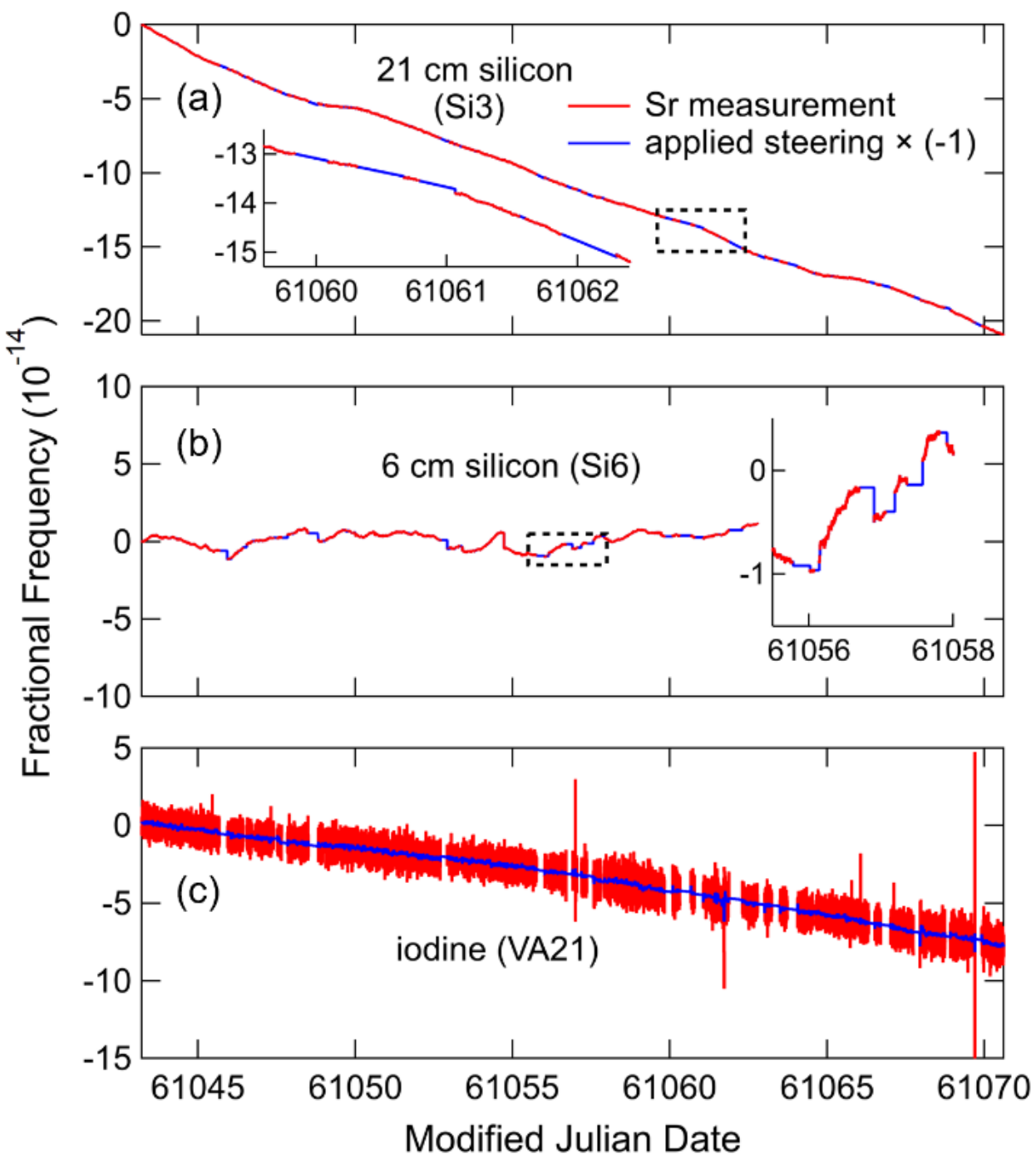

**Figure 3** Steering of the flywheels. (a)-(c) The three optical flywheels are steered based on measurements with the Sr optical frequency standard. Simple linear extrapolation is used for steering the silicon cavities and a Kalman filter algorithm is used for the iodine clock. Applying the frequency corrections results in three all-optical time scales.

Applying the frequency corrections to each optical flywheel oscillator results in three all-optical time scales, which we refer to as TS(Si3), TS(Si6), and TS(VA21). The time differences between pairs of all-optical time scales are shown in Fig. 4(a). When the Sr lattice is operating, the time differences remain steady thanks to its constant supply of steering data. During the gaps between Sr lattice operations, the time scales wander and accumulate relative time differences. The accumulated time difference for each gap is strongly correlated with the duration of the gap and is about ~20 ps for typical ~6 hour gaps (see Methods). Once the gap exceeds 10 hours as in MJD 61062, a relatively large time difference of up to ~100 ps can accumulate. The peak-to-peak time difference over the whole measurement period is ~200 ps, and the final value is <100 ps at the end of the measurement period.

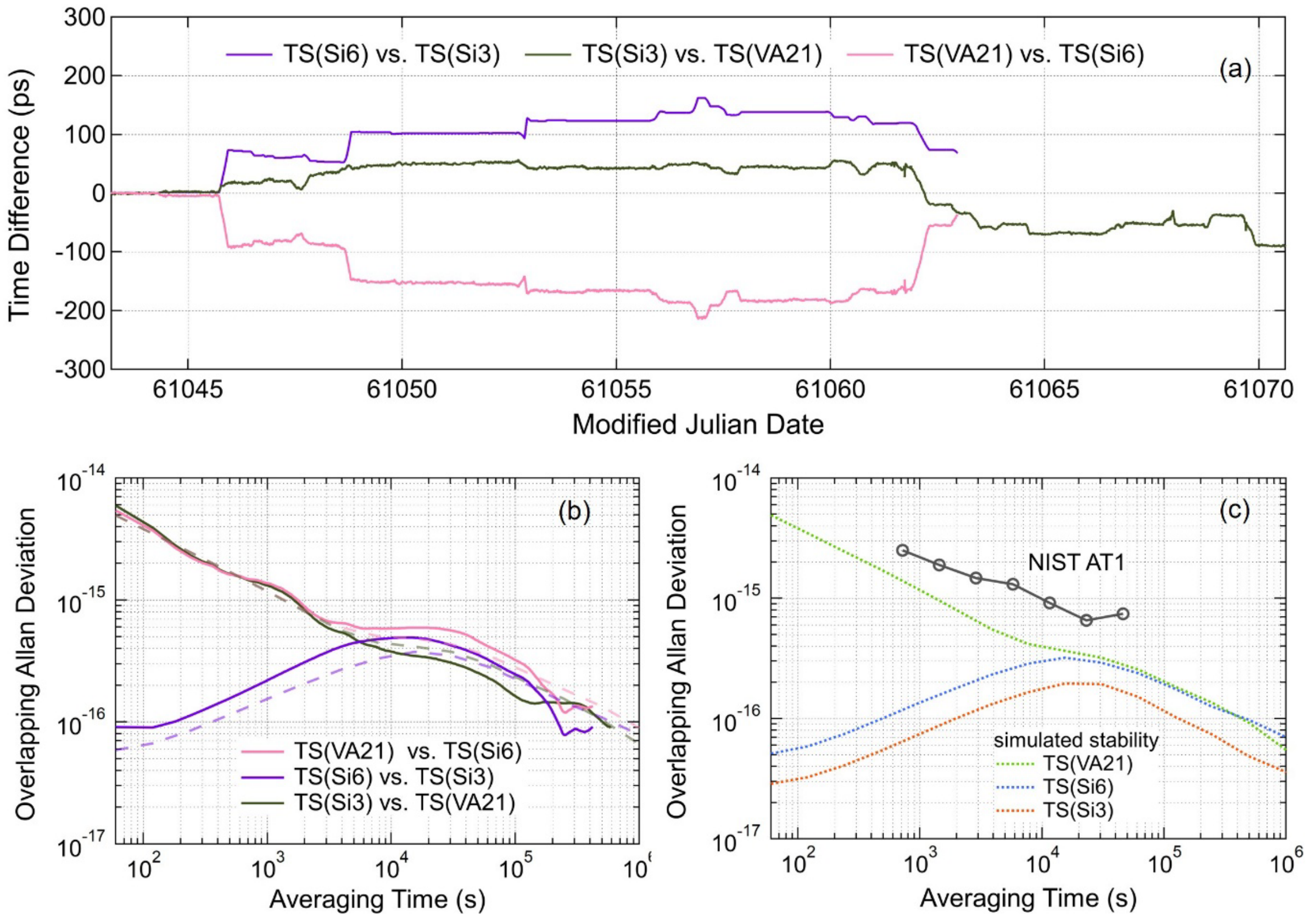

**Figure 4** Comparison of three all-optical time scales. (a) Time differences between pairs of all-optical time scales. (b) Fractional frequency stability of the time scales, computed from the time differences in (a). The dashed lines are simulated relative stability between the time scales based on the noise characteristics of each flywheel and the actual uptime of the Sr frequency reference. (c) Simulated stability of individual all-optical time scales. All three time scales perform better than NIST AT1 time scale based on hydrogen masers.

The relative stability of the time scales, expressed in overlapping Allan deviation, is shown in Fig. 4(b). The measured instability does not exceed mid-$10^{-16}$ at all averaging times, and reaches low-$10^{-16}$ or below after just a few days of averaging. We additionally performed simulations of the all-optical time scales based on the measured noise characteristics of the flywheels and the actual uptime of the Sr lattice (see Methods)[10]. The simulation is able to reproduce the measured relative time scale stability with reasonable agreement (dashed lines). Figure 4(c) shows the simulated stability of the individual all-optical time scales. All three time scales are more stable than the NIST AT1 time scale, which is an ensemble of more than 10 hydrogen masers.

While our all-optical time scales fundamentally reside in the optical domain in the form of laser oscillations, they are phase-coherently converted into 100 MHz rf signals with optical frequency combs. Figure 5(a) shows the frequency difference between two free-running optical flywheels Si3 and Si6, phase-continuously measured in both optical and rf domains (see Extended Data Fig. 5 for a similar continuous record between Si3 and VA21). The rf measurement faithfully follows the optical measurement, albeit with a higher short-term noise coming from the rf measurement device. No artificial offsets or scaling factors are applied other than normalization by respective carrier frequencies. The occasional sharp peaks in the rf measurement are from cycle slips of one of the frequency combs, whose magnitudes are tracked and corrected in post processing (see Methods). The measurement gap in the rf comparison during MJD 61055-61060 is caused by a numerical precision issue from the measurement device[37], and does not imply degradation of the signal quality.

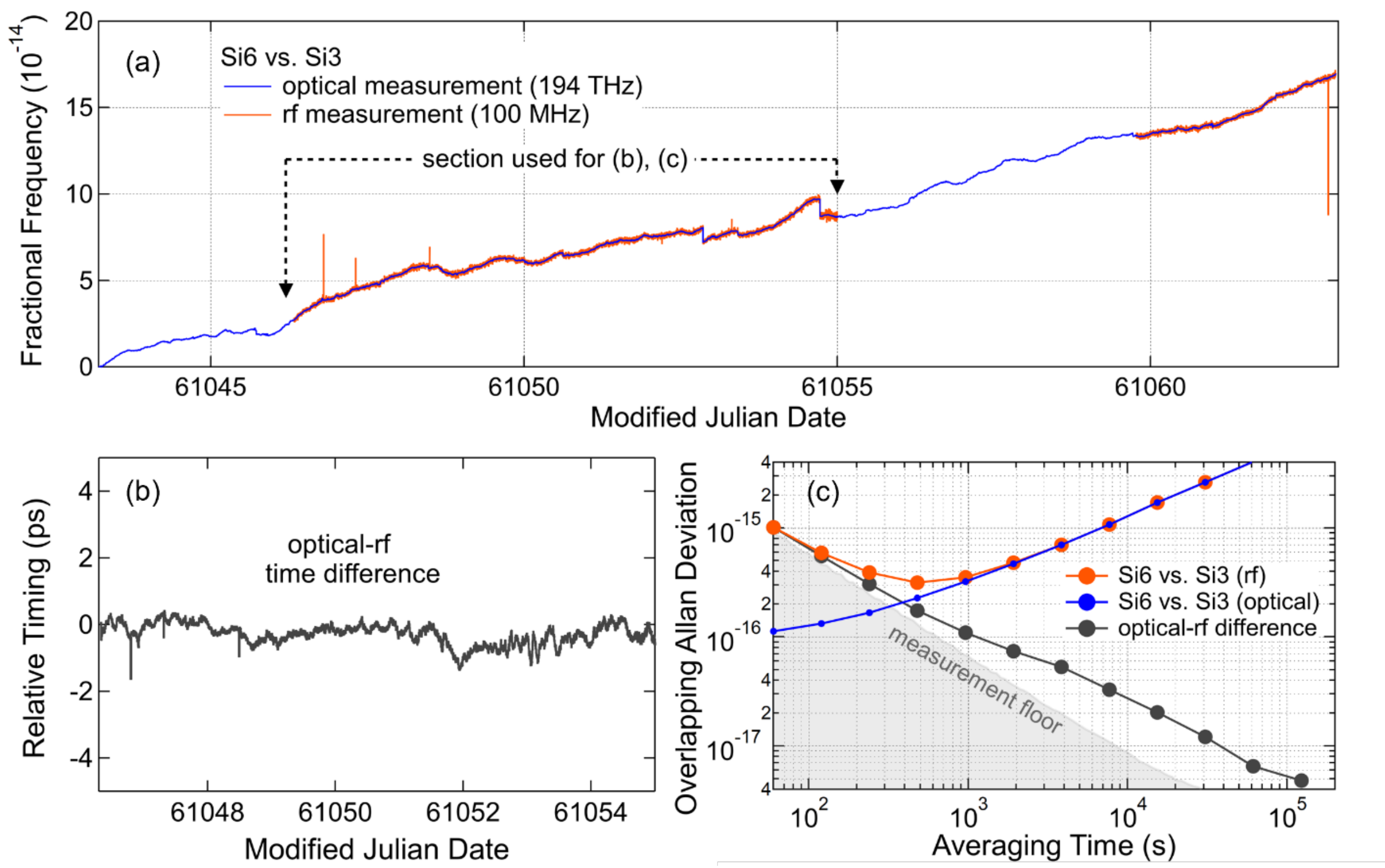

**Figure 5** Fidelity of the optical-to-rf conversion process with an optical frequency comb. (a) Comparison of the Si6-Si3 beat note measured in optical and rf domains. One-minute averaging interval is used. (b) Timing error added in the optical-to-rf conversion process, always showing <2 ps error. (c) Overlapping Allan deviation of the rf and optical beat notes, as well as their difference. Linear drift is not removed. The measurement floor is set by the phase meter (Microchip 53100A).

The fidelity of the optical-to-rf conversion process is quantitatively shown in Fig. 5(b) with the relative timing difference between the optical and rf signals. The residual slope of $(-0.6 \pm 5) \times 10^{-}$

$^{18}$ from a linear fit is consistent with zero, which confirms the accuracy of the frequency division process at that level. The corresponding overlapping Allan deviation of the optical-rf difference, as well as the measured relative stability of the rf signals derived from Si3 and Si6, are limited by the commercial measurement device up to 100 s (Fig. 5(c)). A complete evaluation of the true short-term stability of the rf signals will require a dedicated experiment[38]. Nonetheless, the measurement floor of the device and the $<2$ ps additive noise from the optical-to-rf conversion are low enough to preserve the long-term performance of the all-optical time scales shown in Fig. 4.

**Discussion and conclusion**

Currently, the stability and accuracy of the best time scales are limited to mid-$10^{-16}$ after many days of averaging. With the near-future redefinition of the SI second by an optical frequency, the possibility of keeping time with two orders of magnitude improved performance will emerge. However, time scales with rf flywheels will not be able to realize this level of performance because of the high short-term noise of hydrogen masers. The noise of hydrogen masers requires an unreasonable amount of continuous interrogation to realize the $10^{-18}$-level stability and accuracy of leading optical frequency standards. As demonstrated here, cryogenic silicon cavities and commercially available iodine clocks are ready for improved time scales, enabling near-term adoption of robust optical systems while maintaining compatibility with conventional timebase infrastructure. Cryogenic silicon cavities can in fact fully optimize advanced optical frequency standards. They show $10^{-17}$-level short-term flicker frequency noise limited by fundamental coating Brownian thermal noise[39], and random walk frequency noise at longer terms. Because of the exceptional short-term performance, a silicon cavity steered for one hour per day will still outperform a hydrogen maser steered for 12 hours a day[10]. In addition, the stability of a silicon cavity-based time scale will greatly improve with higher optical frequency standard uptimes, eventually approaching the ultimate limit set by the optical frequency standard under near-continuous operation.

Besides the improved performance, the high carrier frequency of optical flywheels brings several additional advantages. Unlike in phase and frequency metrology of high-performance rf oscillators, the measurement device noise from frequency counters and phase meters is negligible when measuring the residual difference-tones of optical oscillators, even at the state-of-the-art level. For example, a frequency counter with $10^{-12}$-level measurement floor at 1 s can support 10 MHz optical heterodyne beat note measurement with $10^{-19}$-level instability at 1 s. Likewise, the additive noise of optical frequency steering devices such as acousto-optic modulators is negligible for all-optical time scales. Thus, characterization and manipulation of all-optical time scales are much more straightforward.

For all-optical time scales to see widespread use, future work is needed on multiple fronts, all of which are seeing rapid progress. First, reliable and high-uptime optical frequency standards should be widely adopted. The stability and accuracy of these optical frequency standards for time scales need not be at the state-of-the-art level, only better than the $10^{-16}$ level achieved by the best microwave standards. Reliable optical flywheels such as cryogenic silicon cavities should be developed and commercialized with guaranteed performance at a level suitable for a time scale flywheel oscillator. Since intercontinental time and frequency transfer links through satellites support only low-$10^{-16}$ level stability after a few days of averaging, low-noise optical networks need to be deployed for distribution of all-optical time scales in the future[40–42].

In summary, we have realized three all-optical time scales based on three independent optical flywheel oscillators (two cryogenic silicon cavities and one iodine optical clock), all steered to a Sr optical frequency standard. The three all-optical time scales are phase-continuously compared with each other, accumulating <100 ps time difference after more than 20 days of operation and reaching fractional frequency instability below $1 \times 10^{-16}$ after just a few days of averaging. As the SI second is redefined by an optical frequency, the most natural realization of the optical second must employ optical flywheel oscillators, which we show is feasible with significant performance enhancements. With further developments in optical flywheels and optical frequency distribution networks, we expect all-optical time scales to be the future of timekeeping.

# Methods

### Distribution of time difference vs. gap duration

When the Sr lattice is operational, the time scales do not accumulate substantial time difference as the frequency errors from the oscillators are constantly corrected. Large time differences are accumulated only during gaps of the Sr lattice operation. Although the true timing error of individual time scales cannot be measured without a superior time scale, we can measure the time difference accumulated between each pair of time scales during a steering gap. Extended Data Fig. 1(a) shows the distribution of time differences as a function of gap length. As expected, longer gaps cause larger time differences to accumulate. During a gap of length $\tau$, the accumulated time difference is expected to be of order TDEV($\tau$) of the flywheels. We see that this general trend is followed. Extended Data Fig. 1(b) shows the distribution of gap lengths, and we see that most of the gaps are <20 min; brief interruptions due to a laser unlock. However, the gaps over 1 h dominate the accumulated time differences.

### Simulation of all-optical time scales

Extended Data Fig. 2(a) shows the noise model used for simulating the flywheels and corresponding time scales. The model parameters were estimated from the drift-removed overlapping Allan deviations of each flywheel shown in Fig. 2(c). The Si3 model includes a flicker frequency noise of $3.5 \times 10^{-17}$ and a random-walk frequency noise of $3 \times 10^{-18}$ at 1 s. The Si6 model similarly includes a flicker frequency noise of $6 \times 10^{-17}$ and a random walk frequency noise of $6.5 \times 10^{-18}$ at 1 s. For VA21 and its distribution system, a white frequency noise of $4 \times 10^{-14}$ at 1 s was included together with flicker frequency noise of $3 \times 10^{-16}$ and a random walk frequency noise of $2 \times 10^{-18}$ at 1 s. The simulated time domain flywheel records were then steered using the same prediction algorithms used for experimental data analysis. The actual uptime pattern of the Sr lattice, shown in Extended Data Fig. 2(b), was used to mask the simulated flywheel records. We simulated 365 days of flywheel records and repeatedly applied the same Sr uptime pattern to match the simulated data length. The time error of the simulated time scales was determined from the difference between the simulated flywheel frequency and the algorithm prediction, and its stability is shown in Fig. 4(c).

The validity of the simulation was confirmed by reconstructing the relative time differences between pairs of simulated time scales. A comparison with experimentally measured time differences is shown in Fig. 4(b) in the main text. The agreement between the simulation and measurement confirms that the simulated time scales capture all essential features of the experimental time scales.

**Correction of frequency comb cycle slips**

One of the three optical frequency combs employed in the experiment exhibited regular cycle slips on its optical phase lock to Si3. No cycle slips were observed in all other frequency comb locks. If uncorrected, these cycle slips result in relative timing errors between the optical and rf signals, as shown in Extended Data Fig. 3(a). The magnitudes of the cycle slips are tracked and corrected in post processing. To track the number of cycles slipped, we continuously measure the relative phase of the phase-locked beat note between the comb and Si3 with respect to the reference rf signal used for the phase lock, using a zero-deadtime phase meter (K+K FXE phase + frequency meter). When there is no cycle slip, the beat note phase closely follows that of the reference rf signal, as expected from a phase-locked loop. When a cycle slip occurs, the relative phase between the beat note and the reference rf signal jumps by an integer multiple of $2\pi$, as shown in Extended Data Fig. 3(b). These cycle slips are converted to time errors by multiplying by the optical cycle period of 1/194 THz ≈ 5 fs, and are corrected to produce Fig. 5(b) in the main text.

**Generation of 100 MHz signals from silicon cavities**

Two 100 MHz signals were generated from Si3 and Si6 using two independent optical frequency combs with repetition rates of 200 MHz and 250 MHz. The 1 GHz harmonic of the repetition rates was photodetected and frequency-divided to 100 MHz using a module manufactured by Vector Atomic. Extended Data Fig. 4(a) shows the phase noise of the two 100 MHz signals as measured by a Microchip 53100A phase meter. The measured phase noise of -124 dBc/Hz at 1 Hz Fourier frequency is close to the specified measurement floor of the phase meter of -120 dBc/Hz. Extended

Data Fig. 4(b) shows the overlapping Allan deviation of the 100 MHz signals, similar to Fig. 5(c) in the main text. In order to achieve a lower measurement floor by rejecting the input ADC noise, cross-correlation overlapping Allan deviation was used[43]. The measured 1 s stability of $6.7 \times 10^{-15}$ is still close to the specified measurement floor of $7 \times 10^{-15}$.

**Acknowledgements**

Funding for this work is provided by NSF No. QLCI OMA-2016244, Defense Advanced Research Projects Agency contract HR00112590168 (OASIC), V. Bush Fellowship, Sloan and Simons Foundaton, NSF No. PHY-2317149, and NIST.

**Author Contributions**

The experimental demonstration of the optical time scale was carried out in the JILA laboratory with all authors contributing to the overall operation. All authors contributed to the data analysis and writing of the manuscript, led by D.L.

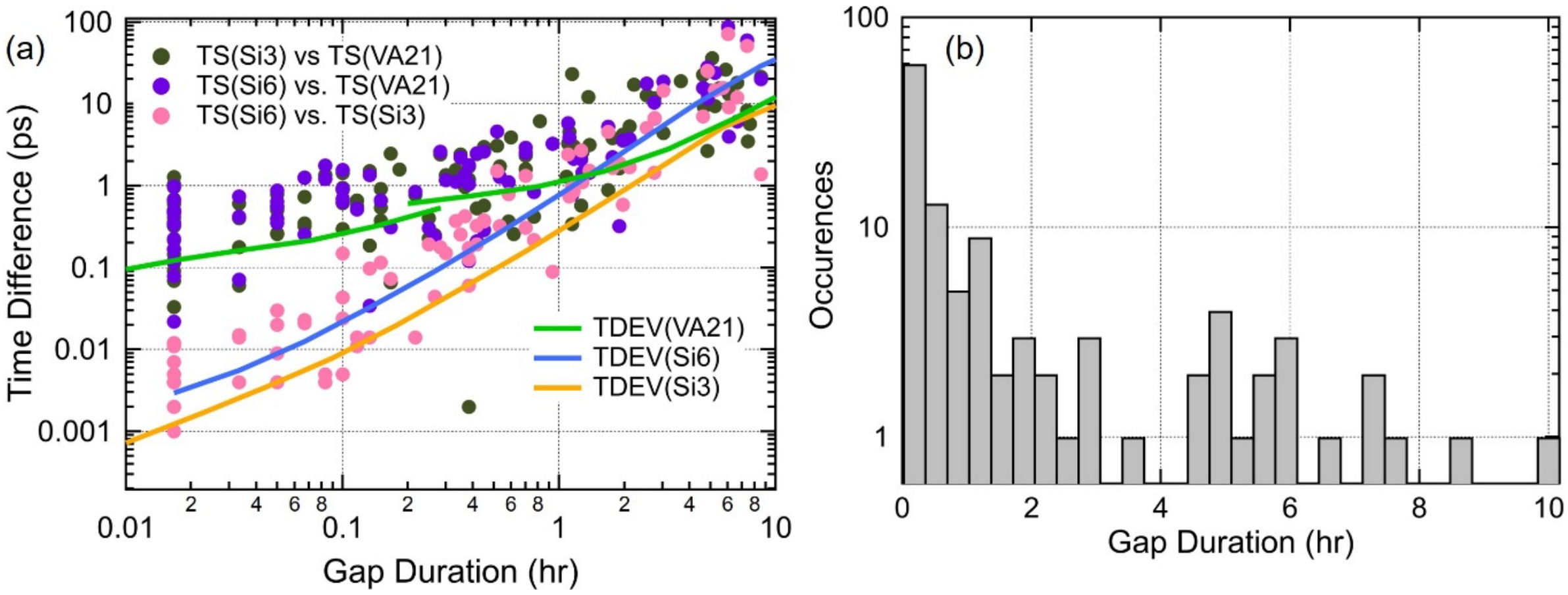

**Extended Data Figure 1** Distribution of time difference vs. gap duration. (a) The time difference accumulated between each pair of time scales during gaps of different length. The expected time difference during the gaps is given by the time deviation of each flywheel, shown as solid lines. (b) A histogram showing the distribution of gap lengths in the Sr lattice operation.

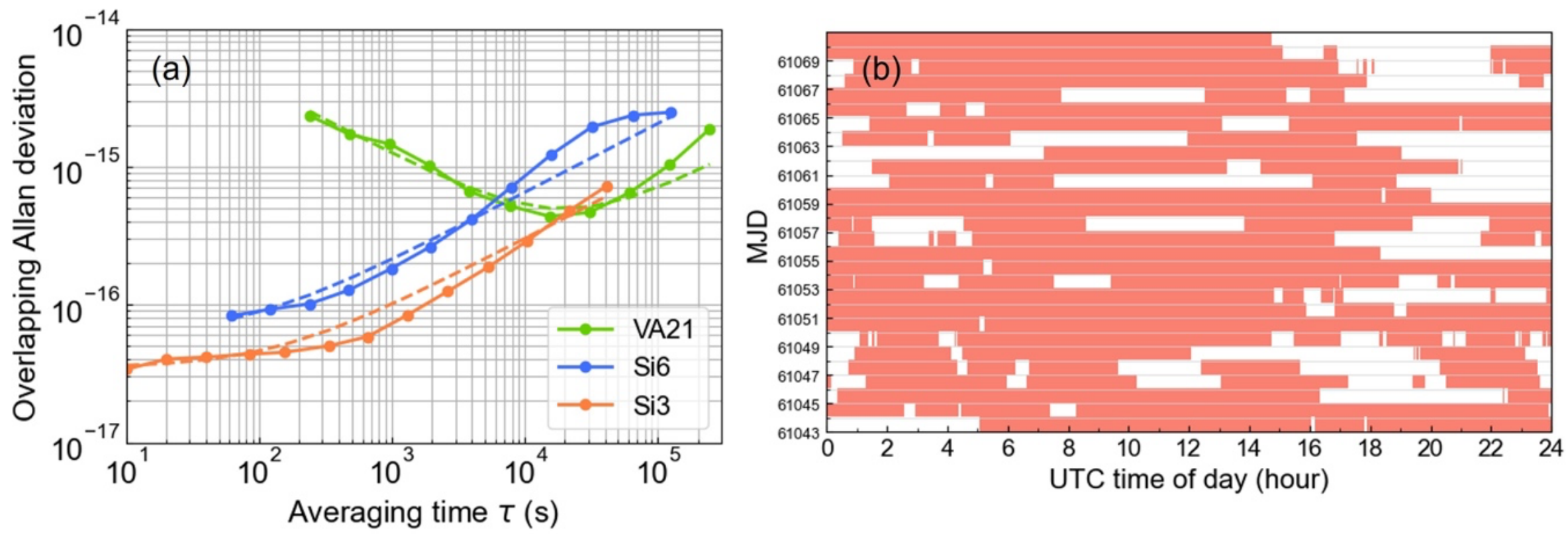

**Extended Data Figure 2** Flywheel noise models and Sr uptime. (a) Noise models (dashed lines) determined from the overlapping Allan deviations of each flywheel oscillator (solid marked lines). (b) Uptime of the Sr lattice during the measurement period. Red shading indicates periods of Sr operation. Uptime is reduced during working hours (UTC 1600 to 0100).

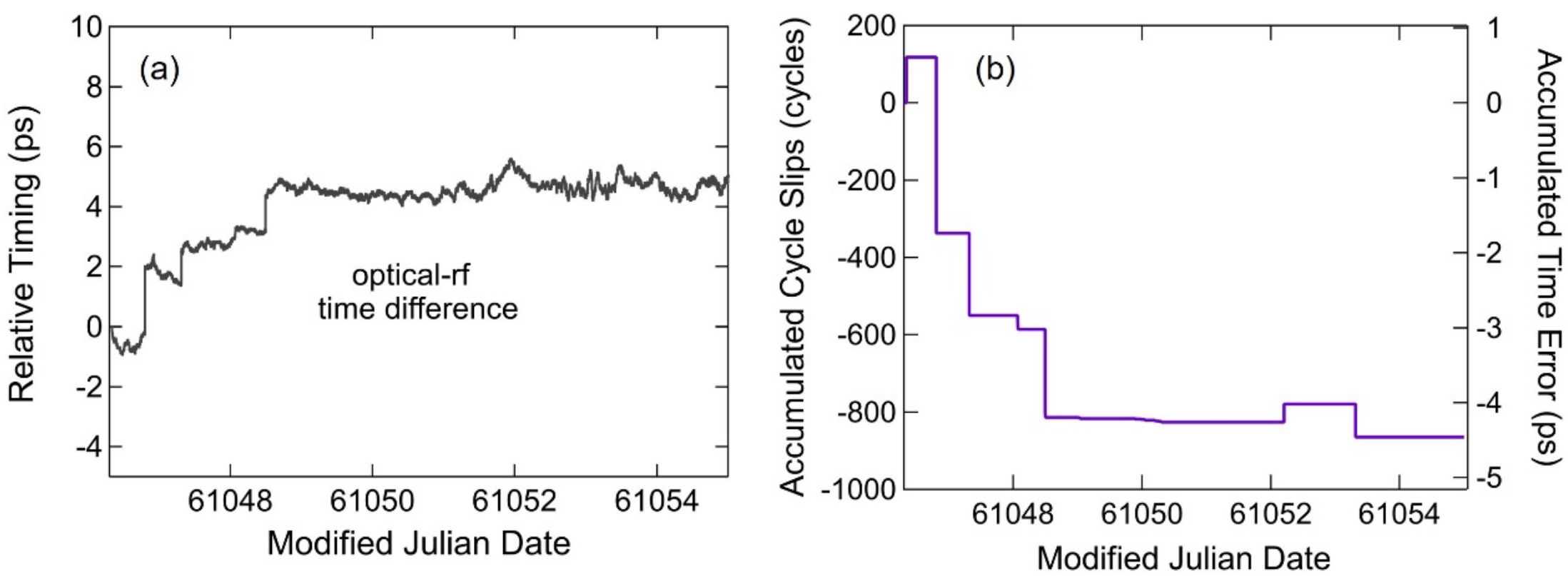

**Extended Data Figure 3** Correction of frequency comb cycle slips. (a) Timing error occurred in the optical-to-rf conversion process before correcting for cycle slips of the comb. The cycle slips manifest as discrete jumps in the relative timing between optical and rf signals. (b) Accumulated cycle slips on the comb-Si3 phase locked loop. The number of cycle slips is converted to time error by multiplying it by the optical cycle period of ~5 fs (right axis). These independently recorded cycle slips are used to correct timing errors.

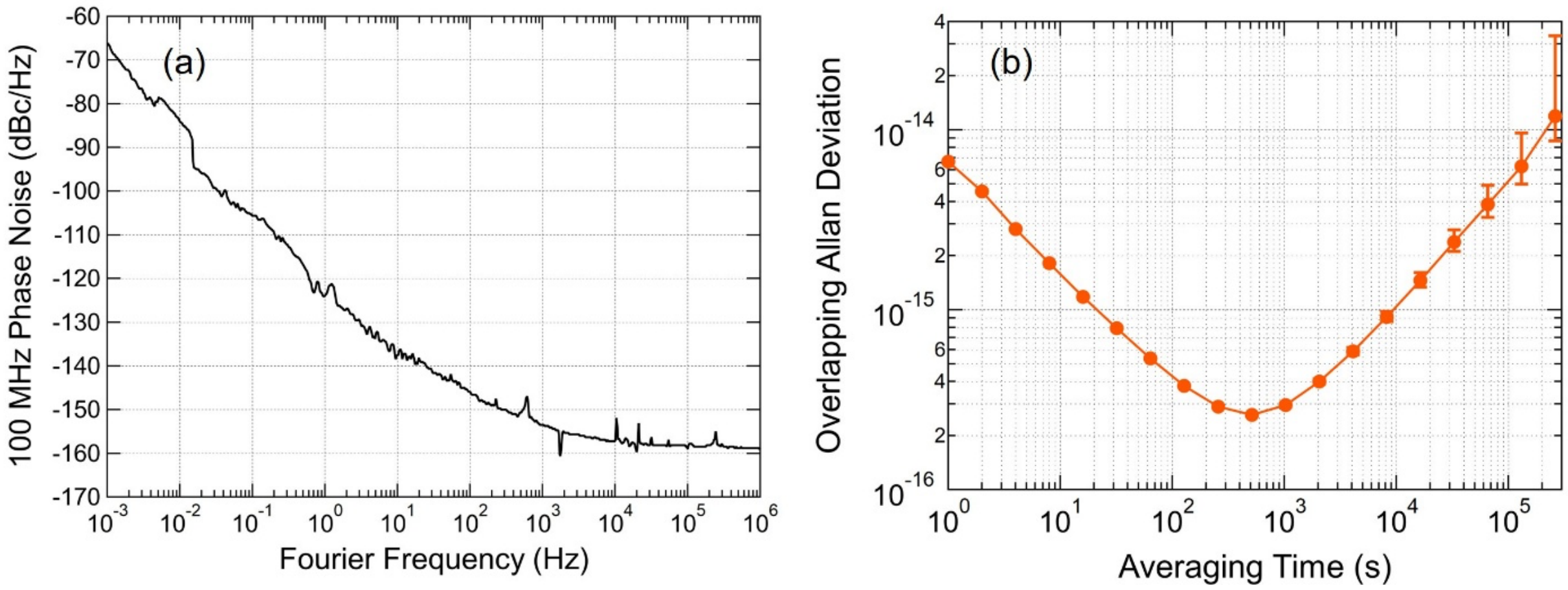

**Extended Data Figure 4** Relative noise of two 100 MHz signals derived from their respective silicon cavities. Neither cavity is steered by the Sr atoms. (a) Phase noise of the 100 MHz signals, measured with a Microchip 53100A phase meter. The plotted data include contributions from both 100 MHz signals. At 1 Hz Fourier frequency, the device-specified noise floor sets the measurement limit. (b) Frequency stability of the 100 MHz signals. Cross correlation method was used to achieve a lower measurement floor. The measured stability at 1 s is dominated by the measurement device noise.

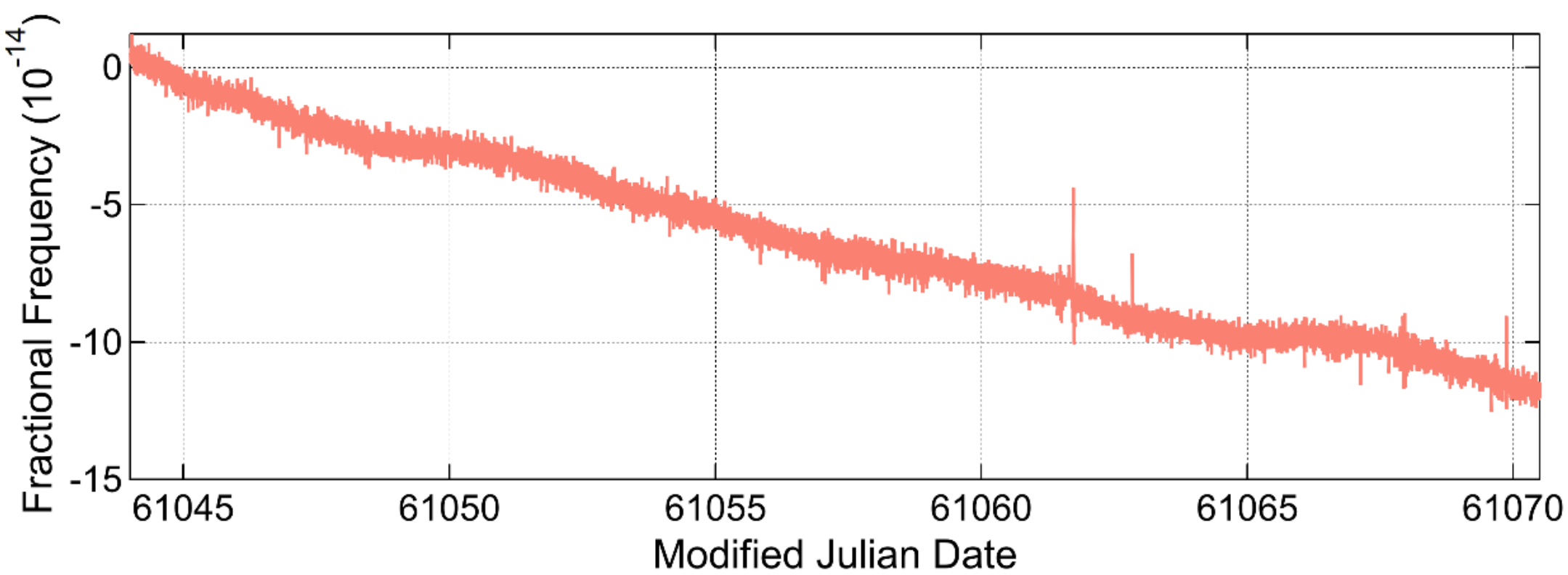

**Extended Data Figure 5** Analogous to the continuous comparison between Si3 and Si6 in Fig. 5(a), the relative frequency and phase of Si3 (10 MHz) and VA21 (5 MHz) are measured with 100% uptime. 10-minute averaging interval is used. The occasional spikes are from the fiber link between JILA and NIST.